\UseRawInputEncoding
\documentclass[pra,superscriptaddress]{revtex4}
\usepackage{graphicx}
\usepackage{dcolumn}
\usepackage{bm}
\usepackage{amsmath}
\usepackage{amssymb,amsbsy,cancel}
\usepackage{color,ulem}
\usepackage{soul}
\newcommand{\beq}{\begin{equation}}
\newcommand{\eeq}{\end{equation}}
\newcommand{\bea}{\begin{eqnarray}}
\newcommand{\eea}{\end{eqnarray}}
\newcommand{\bec}{\begin{center}}
\newcommand{\enc}{\end{center}}
\newcommand{\bfr}{\begin{flushright}}
\newcommand{\efr}{\end{flushright}}
\newcommand{\la}{\langle}
\newcommand{\ra}{\rangle}

\newcommand{\alp}{\alpha}

\newcommand{\om}{\omega}
\newcommand{\kap}{\kappa}
\newcommand{\gam}{\gamma}
\newcommand{\sig}{\sigma}

\newcommand{\tone}{\widetilde{1}}
\newcommand{\ttwo}{\widetilde{2}}
\newcommand{\tthree}{\widetilde{3}}
\newcommand{\tfour}{\widetilde{4}}

\newcommand{\ha}{\hat{a}}

\newcommand{\hsig}{\hat{\sigma}}
\newcommand{\mra}{\mathrm{a}}
\newcommand{\mrc}{\mathrm{c}}

\newcommand{\mrp}{\mathrm{p}}
\newcommand{\mrs}{\mathrm{s}}
\begin{document}
\title{
Demonstration of deterministic SWAP gate between 
superconducting and frequency-encoded microwave-photon qubits
}
\author{Kazuki Koshino}
\email{kazuki.koshino@osamember.org}
\affiliation{College of Liberal Arts and Sciences, Tokyo Medical and Dental
University, Ichikawa, Chiba 272-0827, Japan}
\author{Kunihiro Inomata}
\email{kunihiro.inomata@aist.go.jp}
\affiliation{National Institute of Advanced Industrial Science 
and Technology (AIST), 
Tsukuba, Ibaraki 305-8568, Japan}
\date{\today}
\begin{abstract}
The number of superconducting qubits contained in a single quantum processor is increasing steadily. 
However, to realize a truly useful quantum computer, 
it is inevitable to increase the number of qubits much further
by distributing quantum information among distant processors using flying qubits. 
Here, we demonstrate a key element towards this goal, namely, 
a SWAP gate between the superconducting-atom and microwave-photon qubits. 
The working principle of this gate is the single-photon Raman interaction, 
which results from strong interference in one-dimensional optical systems 
and enables a high gate fidelity insensitively to the pulse shape of the photon qubit, 
by simply bouncing the photon qubit at a cavity attached to the atom qubit. 
We confirm the bidirectional quantum state transfer between the atom and photon qubits. 
The averaged fidelity of the photon-to-atom (atom-to-photon) state transfer reaches 0.829 (0.801), 
limited mainly by the energy relaxation time of the atom qubit. 
The present atom-photon gate, equipped with an {\it in situ} tunability of the gate type, 
would enable various applications in distributed quantum computation using superconducting qubits and microwave photons.
\end{abstract}
\maketitle

\section{introduction}
The number of solid-state qubits contained in a single processor 
is steadily increasing~\cite{qc2,qc3} and has reached 3 digits recently. 
However, an incomparably larger number of qubits is required
in order to make such quantum machines truly useful. 
Therefore, in near future, 
it would be indispensable to distribute quantum information 
among remote quantum processors~\cite{dist1,dist2,dist3,dist4}, 
using deterministic quantum interactions 
between stationary and flying qubits~\cite{atom_photon_1,atom_photon_2}.

In superconducting quantum computation, 
stationary qubits are encoded on various types of superconducting atoms
and flying qubits are encoded on microwave photons propagating in waveguides. 
In such setups, the atom-photon interaction is drastically enhanced
owing to the natural spatial mode-matching between radiation from the atom
and a propagating photon in the waveguide~\cite{MWqo1,MWqo2,MWqo3,MWqo4,MWqo5,MWqo6,MWqo7}. 
Applying such waveguide QED effects, 
single microwave photon detection has been accomplished~\cite{spd1,spd2,spd3,spd4,spd5,spd6}. 
Another prominent achievement in this field is 
the deterministic release and catch of a photon by remote atoms, 
which is accomplished by tuning the atom-waveguide coupling 
and has been applied for remote entanglement generation 
and a photon-photon gate~\cite{catch_0,catch_1,catch_2,catch_3,catch_4,catch_5,catch_6}. 
A technical difficulty with such {\it active} atom-photon interaction  
is the need for precise temporal control of the atom-waveguide coupling
in accordance with the pulse shape of the photon, 
without which the capture probability of propagating photons is substantially lost.

In this paper, we demonstrate a deterministic SWAP gate 
between a superconducting atom and a microwave photon~\cite{prop}. 
This gate is based on an essentially {\it passive} working principle, 
the single-photon Raman interaction~\cite{dop_th_1,dop_th_2,dop_th_3,our_1,our_2,dayan1,dayan2,dayan3}.
This is a phenomenon characteristic to one-dimensional optical systems, 
originating in the strong destructive interference 
between an applied photon field to an emitter and radiation from it.  
This guarantees a high-fidelity gate operation
insensitively to the shape and length of the input photon qubit.
Besides this point, 
the present scheme has the following merits for practical implementation. 
Simple setup---the required system for the present gate 
is an atom and a resonator coupled in the dispersive regime, 
each coupled to independent waveguides (Fig.~\ref{fig:setup}). 
This is a quite common element in a superconducting quantum processor
adopting the dispersive qubit readout~\cite{dr1,dr2}.
Gate tunability---although we demonstrate only the SWAP gate in this work, 
the present gate is a more general $(\mathrm{SWAP})^{\alp}$ gate 
($0 \leq \alp \leq 1$)~\cite{swap_a_1,swap_a_2,swap_a_3,swap_a_4}
equipped with an {\it in situ} tunability of the gate type $\alp$
through the amplitude and frequency of the drive pulse to the atom~\cite{prop}. 
In particular, in combination with the atom-photon entangling gate
($\sqrt{\mathrm{SWAP}}$, $\alp=1/2$), 
the present scheme is applicable to 
the entanglement generation between remote superconducting qubits
as well as the deterministic photon-photon entangling gate.
Dual-rail encoding of photon qubit---the photon qubit 
is encoded on its two different carrier frequencies~\cite{freq0,freq1,freq2,freq3}. 
In contrast with the single-rail (photon number) encoding, 
we can avoid degradation of fidelity by photon loss
and sharing a phase reference~\cite{jesper}.

The rest of this paper is organized as follows.
In Sec.~\ref{sec:gate}, we present the setup to realize the SWAP gate 
between a superconducting atom and a microwave photon
and explain the working principle of the gate. 
In Sec.~\ref{sec:demo}, we demonstrate the atom-photon SWAP gate.
More concretely, we confirm both photon-to-atom and atom-to-photon qubit transfer
by constructing the final atom/photon density matrix.
The photon qubit in the present gate is a single photon in principle,
but we use a weak coherent-state photon instead in this work. 
Section~\ref{sec:sum} is devoted to summary. 
In Appendices~\ref{app:PtoA} and \ref{app:AtoP}, details on the density matrix construction are presented. 
In Appendix~\ref{app:exp}, details on the experimental information are presented. 

\section{Atom-photon SWAP gate}\label{sec:gate}
\subsection{Setup}
\begin{figure*}[t]
\begin{center}
\includegraphics[width=150mm]{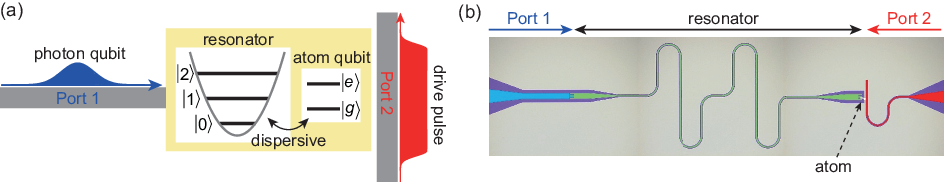}
\end{center}
\caption{
Setup for the SWAP gate between superconducting-atom and microwave-photon qubits.
(a)~Schematic of the setup. 
(b)~False-colored optical micrograph of the actual device.
}
\label{fig:setup}
\end{figure*}
The setup for the present atom-photon SWAP gate is 
a common one in superconducting quantum computing: 
a superconducting atom is dispersively coupled to a microwave resonator, 
and transmission lines are attached to both of them (Fig.~\ref{fig:setup}). 
One of the lines (Port~2) is coupled to the atom and a microwave drive pulse, 
which transforms the {\it bare} states of the atom-resonator system to 
the {\it dressed} ones within the pulse duration, 
is applied through this line. 
The other line (Port~1) is coupled to the resonator
and a single microwave photon, which serves as a photon qubit, 
is input through this line synchronously with the drive pulse. 
The atom qubit is encoded on its ground and excited states, $|g\ra$ and $|e\ra$. 
The photon qubit is encoded on its two different carrier frequencies, 
$|\om_L\ra$ and $|\om_H\ra$, 
where $\om_L$ ($\om_H$) denotes the lower (higher) carrier frequency.
The gate operation completes deterministically 
by bouncing the photon qubit applied through Port~1.  

\subsection{Principle of SWAP gate}
\begin{figure}[t]
\begin{center}
\includegraphics[width=70mm]{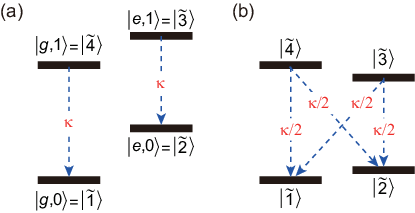}
\end{center}
\caption{
Level structure of the atom-resonator system.
(a)~Bare states, where the drive field from Port~2 is off.
(b)~Dressed states, where the drive field is on. 
Note that the energy diagram (b) is drawn 
in the frame rotating at the drive frequency $\om_\mathrm{d}$. 
}
\label{fig:lvs}
\end{figure}
Here, we outline the working principle of the present gate. 
We label the eigenstates of the atom-resonator system by $|a,n\ra$, 
where $a=\{g, e\}$ and $n=\{0, 1, \cdots\}$ respectively specify
the atomic state and the photon number in the resonator. 
The atom-resonator system is in the dispersive coupling regime, 
and the eigenfrequencies are given by
$\om_{|g,n\ra}=n\om_\mathrm{r}$ and 
$\om_{|e,n\ra}=\om_\mathrm{ge}+n(\om_\mathrm{r}-2\chi)$, 
where $\om_\mathrm{ge}$ and $\om_\mathrm{r}$ are the renormalized 
frequencies of the atom and the resonator and $\chi$ is the dispersive shift.
In the present atom-photon gate, 
we use the lowest four levels of the atom-resonator system
($a=\{g, e\}$ and $n=\{0, 1\}$). 
The principal decay channel of this four-level system 
is the radiative decay of the resonator to Port~1 with a rate of $\kappa$. 
Therefore, when the drive field is off, 
the radiative decay within this four-level system occurs vertically with $\kappa$
[Fig.~\ref{fig:lvs}(a)].

During the interaction between the photon qubit and the atom-resonator system, 
we apply a microwave drive to the atom from Port~2.  
This drive field hybridizes the lower-two {\it bare} states $|g,0\ra$ and $|e,0\ra$ 
to form the {\it dressed} states $|\tone\ra$ and $|\ttwo\ra$. 
By switching on/off the drive field adiabatically, 
we can convert the bare and dressed states deterministically as
$|g,0\ra \leftrightarrow |\tone\ra$ and 
$|e,0\ra \leftrightarrow |\ttwo\ra$. 
Similarly, the higher-two states are converted as
$|e,1\ra \leftrightarrow |\tthree\ra$, and 
$|g,1\ra \leftrightarrow |\tfour\ra$.   
In addition to vertical decays ($|\tthree\ra \to |\ttwo\ra$ and $|\tfour\ra \to |\tone\ra$), 
oblique decays ($|\tthree\ra \to |\tone\ra$ and $|\tfour\ra \to |\ttwo\ra$) 
become allowed due to hybridization in a dressed-state basis.

In particular, under a proper choice of the frequency and power of the drive field, 
the four radiative decay rates take an identical value of $\kap/2$  [Fig.~\ref{fig:lvs}(b)].
Then, the levels $|\tone\ra$, $|\ttwo\ra$ and $|\widetilde{j}\ra$ ($j=3$ or 4) 
function as an ^^ ^^ impedance-matched'' $\Lambda$ system~\cite{prop,our_1}:
if the system is in the state $|\tone\ra$ initially and a single photon 
with frequency $\om_{j1}=\om_{|\widetilde{j}\ra}-\om_{|\widetilde{1}\ra}$ is input from Port~1, 
a Raman transition $|\tone\ra \to |\widetilde{j}\ra \to |\ttwo\ra$ is deterministically induced.
As a result, the $\Lambda$ system switches to the state $|\ttwo\ra$ and 
the input photon becomes down-converted to frequency $\om_{j2}$ after reflection. 
In this study, we choose $j=4$ and set $\om_L=\om_{42}$ and $\om_H=\om_{41}$ 
as the lower and higher carrier frequencies of the photon qubit. 
The time evolution of the atom and photon qubits is then written as
$|\tone, \om_H\ra \to |\ttwo, \om_L\ra$.
The inverse process, $|\ttwo, \om_L\ra \to |\tone, \om_H\ra$, is also deterministic. 
In contrast, for the initial states of $|\tone,\om_L\ra$ and $|\ttwo,\om_H\ra$, 
the input photon is perfectly reflected as it is without interacting with the $\Lambda$ system, 
since the input photon is out of resonance of the $\Lambda$ system. 
Namely, $|\tone,\om_L\ra \to |\tone,\om_L\ra$ and $|\ttwo,\om_H\ra \to |\ttwo,\om_H\ra$. 
These four-time evolutions are summarized as
\bea
(\beta_1|\tone\ra + \beta_2|\ttwo\ra)\otimes(\gam_1|\om_L\ra + \gam_2|\om_H\ra)
\to
(\gam_1|\tone\ra + \gam_2|\ttwo\ra)\otimes(\beta_1|\om_L\ra + \beta_2|\om_H\ra),
\label{eq:swap}
\eea
where $\beta_1, \cdots, \gam_2$ are arbitrary coefficients
satisfying $|\beta_1|^2+|\beta_2|^2=|\gam_1|^2+|\gam_2|^2=1$.

Before and after the interaction between the photon qubit and the atom-resonator system,
we switch off the drive field. 
Therefore, the atom-resonator system returns to the bare state basis as
$|\tone\ra =|g,0\ra$ and $|\ttwo\ra =|e,0\ra$. 
Omitting the resonator's state, 
which is in the vacuum state at both the initial and final moments, 
Eq.~(\ref{eq:swap}) is rewritten as
\bea
(\beta_1|g\ra + \beta_2|e\ra)\otimes(\gam_1|\om_L\ra + \gam_2|\om_H\ra)
\to
(\gam_1|g\ra + \gam_2|e\ra)\otimes(\beta_1|\om_L\ra + \beta_2|\om_H\ra).
\label{eq:swap2}
\eea
This is a SWAP gate between the atom qubit and 
the photon qubit applied through Port~1.

\section{demonstration of SWAP gate}
\label{sec:demo}
In this section, we demonstrate the atom-photon SWAP gate. 
This should be done, in principle, by applying a single-photon pulse from Port~1 as the photon qubit. 
However, in this study, we use a weak coherent-state pulse instead,
the mean photon number $|\alp|^2$ of which is much smaller than unity.  
This pulse is dichromatic in general 
with the carrier frequencies $\om_L$ and $\om_H$ 
and has a Gaussian temporal profile 
with the pulse length of $t_\mathrm{p}=100$~ns. 
This is long enough to satisfy the condition for high-fidelity atom-photon gate, $t_\mathrm{p} \gg 1/\kap$, 
where $\kap$ is the resonator decay rate into Port~1 (see Table~\ref{tab:params}).
If the initial states of the atom and photon qubits are
$\beta_1|g\ra + \beta_2|e\ra$ and $\gam_1|\om_L\ra + \gam_2|\om_H\ra$, respectively,
the initial state vector of the atom-photon system is written as
\bea
|\psi_i\ra &=& 
(\beta_1|g\ra + \beta_2|e\ra) \otimes
e^{-|\alp|^2/2}\left[
|0\ra + \alp(\gam_1|\om_L\ra + \gam_2|\om_H\ra) + \cdots
\right],
\label{eq:psii}
\eea
where $\alp$ represents the complex amplitude of the input photon-qubit pulse,
and the dots represent the multiphoton components in the pulse,
which are negligible when $|\alp|^2 \ll 1$. 
This state vector is rewritten as 
\bea
|\psi_i\ra &=& 
c_1|g,0\ra + c_2|e,0\ra + c_3|g,\om_L\ra + c_4|e,\om_L\ra + c_5|g,\om_H\ra + c_6|e,\om_H\ra +\cdots,
\label{eq:psii2}
\eea
where $(c_1,\cdots,c_6)=e^{-|\alp|^2/2}\times(\beta_1, \beta_2, \alp\beta_1\gam_1, 
\alp\beta_2\gam_1, \alp\beta_1\gam_2, \alp\beta_2\gam_2)$.

The atom-photon SWAP gate is completed
by bouncing the photon qubit at the capacitance connecting Port~1 and the resonator.
The time evolution is given by Eq.~(\ref{eq:swap})
and results in the following final state vector, 
\bea
|\psi_f\ra &=& 
c_1|g,0\ra + c_2|e,0\ra + c_3|g,\om_L\ra + c_5|e,\om_L\ra + c_4|g,\om_H\ra + c_6|e,\om_H\ra +\cdots.
\label{eq:psif}
\eea

\subsection{State transfer: photon to atom}
\subsubsection{Procedures for density matrix estimation}
In order to demonstrate the atom-photon SWAP gate, 
we confirm the bidirectional state transfer between the atom and photon qubits.
Here, we demonstrate the photon-to-atom state transfer. 
We denote the density matrix of the final atom qubit by $\hat{\rho}^{(\mrc)}_\mra$, 
where the superscript (c) implies that 
the input photon-qubit pulse is in a coherent state. 
The matrix element of $\hat{\rho}^{(\mrc)}_\mra$ is given by
$\hat{\rho}^{(\mrc)}_{\mra,mn}=\mathrm{Tr}_\mrp
\{\la m|\psi_f\ra \la\psi_f|n\ra\}$, 
where $m,n(=g,e)$ specify the atomic state and 
$\mathrm{Tr}_\mrp$ takes the trace over the photonic states.
From Eq.~(\ref{eq:psif}), 
$\rho^\mathrm{(c)}_{\mra,ee}$ and $\rho^\mathrm{(c)}_{\mra,eg}$ are given, 
up to the second order in $|\alp|$, by 
\bea
\rho^{(\mrc)}_{\mra,ee}
& \approx & |\beta_2|^2 + |\alp|^2(|\gam_2|^2-|\beta_2|^2),
\label{eq:xiee}
\\
\rho^{(\mrc)}_{\mra,eg}
& \approx & \beta_1^*\beta_2 + |\alp|^2(\gam_1^*\gam_2-\beta_1^*\beta_2).
\label{eq:xieg}
\eea
On the other hand, our target quantity here is 
the density matrix $\hat{\rho}^{(\mrs)}_\mra$ 
of the final atom qubit assuming the single-photon input. 
If the SWAP gate is performed with the single-photon input, 
the final atomic state is $\gam_1|g\ra + \gam_2|e\ra$ [see Eq.~(\ref{eq:swap2})]. 
Therefore, $\rho^{(\mrs)}_{\mra,ee}=|\gam_2|^2$ 
and $\rho^{(\mrs)}_{\mra,eg}=\gam_1^*\gam_2$.

Thus, we can reproduce the target density matrix $\hat{\rho}^\mathrm{(s)}_\mra$
from the measurable one $\hat{\rho}^{(\mrc)}_\mra$ by the following procedures: 
Setting the initial atomic state at $|g\ra$ [namely, $(\beta_1, \beta_2)=(1,0)$],
we perform the atom-photon SWAP gate and measure 
the final atomic density matrix elements $\rho^{(\mrc)}_{\mra,ee}$ and $\rho^{(\mrc)}_{\mra,eg}$.
Putting $(\beta_1, \beta_2)=(1,0)$ in Eqs.~(\ref{eq:xiee}) and (\ref{eq:xieg}), 
they are expected to behave as
$\rho^{(\mrc)}_{\mra,ee}=|\alp|^2 |\gam_2|^2$ and 
$\rho^{(\mrc)}_{\mra,eg}=|\alp|^2 \gam_1^*\gam_2$. 
Therefore, by varying the mean photon number $|\alp|^2$ 
and measuring the slopes of these quantities, 
two of the target density matrix elements are estimated as
\bea
\rho^{(\mrs)}_{\mra,ee} &=& \frac{d}{d|\alp|^2} \rho^{(\mrc)}_{\mra,ee},
\label{eq:drhoda1}
\\
\rho^{(\mrs)}_{\mra,eg} &=& \frac{d}{d|\alp|^2} \rho^{(\mrc)}_{\mra,eg}. 
\label{eq:drhoda2}
\eea
The other elements are determined by
$\rho^{(\mrs)}_{\mra,ge} = (\rho^{(\mrs)}_{\mra,eg})^*$ and 
$\rho^{(\mrs)}_{\mra,gg} = 1- \rho^{(\mrs)}_{\mra,ee}$.

\begin{figure}[h]
\begin{center}
\includegraphics[scale=1.0]{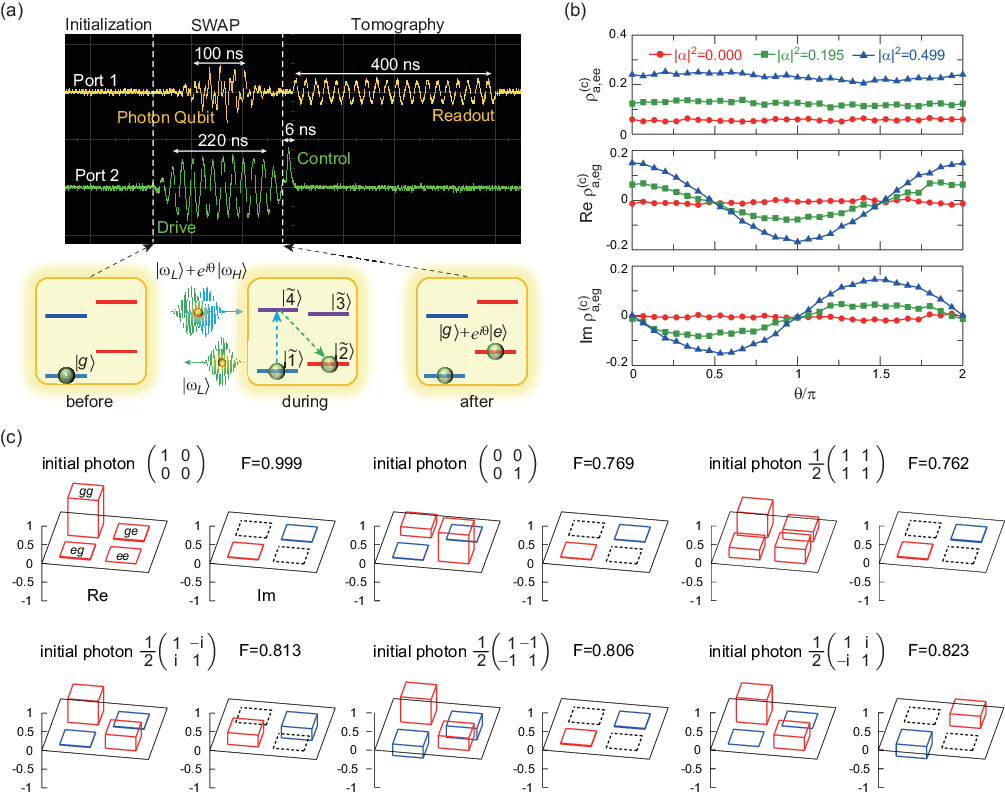}
\end{center}
\caption{
Photon-to-atom state transfer.
(a)~Pulse sequence depicted in the intermediate frequencies (IFs)
and the energy diagrams before, during, and after the SWAP gate.
In the energy diagrams, 
the initial atom (photon) qubit is assumed to be in the polar (equator) state, 
namely, $(\beta_1, \beta_2)=(1, 0)$ and 
$(\gam_1, \gam_2)=(1, e^{i\theta})/\sqrt{2}$.
(b)~Measured density matrix elements
($\rho^{(\mrc)}_{\mra,ee}$, $\mathrm{Re}\rho^{(\mrc)}_{\mra,eg}$, 
$\mathrm{Im}\rho^{(\mrc)}_{\mra,eg}$) 
of the final atom qubit for the coherent-state input.
The initial atom (photon) qubit is in the ground (equator) state, 
and the phase $\theta$ of the photon qubit is varied continuously. 
The mean photon number $|\alp|^2$ in the photon-qubit pulse is indicated.
(c)~Estimated density matrix $\hat{\rho}_{\mra}^{(\mrs)}$
of the final atom qubit assuming the single-photon input.
Positive, negative, and zero matrix elements are drawn in
red, blue, and black dotted lines, respectively.
The fidelity to the initial photon qubit is indicated. 
}
\label{fig:PtoA}
\end{figure}

\subsubsection{Pulse sequence}
The pulse sequence to measure the density matrix $\hat{\rho}^{(\mrc)}_\mra$
of the final atom qubit is shown in Fig.~\ref{fig:PtoA}(a). 
The measurement is composed of three steps. 
(i)~Initialization: we wait for complete de-excitation of the atom applying no pulses. 
(ii)~SWAP gate: from Port~2, we apply a drive pulse with a flat-top envelope at $\omega_{\rm d}/2\pi = 5.785$~GHz to the atom
to implement an {\it impedance-matched} $\Lambda$ system [Fig.~\ref{fig:lvs}(b)]. 
Within the drive pulse duration, we input from Port~1 the photon-qubit pulse with a Gaussian envelope,
which is dichromatic ($\om_L/2\pi=\om_{42}/2\pi = 10.208$~GHz and $\om_H/2\pi=\om_{41}/2\pi = 10.266$~GHz) in general. 
Note that the photon-qubit pulse in Fig.~\ref{fig:PtoA}(a) does not have 
a clear Gaussian envelope because of its dichromatic nature.
(iii)~Tomography: we first apply a short control pulse 
(no pulse, $\pi$ pulse, or 4 kinds of $\pi/2$ pulse)
with a Gaussian envelope at $\omega_{\rm ge}/2\pi = 5.839$~GHz to the atom 
and then dispersively read out the atomic state with a rectangular pulse at $\omega_{\rm r}/2\pi = 10.258$~GHz.

\subsubsection{Results and discussion}
The measured density matrix elements $\rho^{(\mrc)}_{\mra,ee}$ and 
$\rho^{(\mrc)}_{\mra,eg}$ of the final atom qubit are shown in Fig.~\ref{fig:PtoA}(b), 
where the initial atom (photon) qubit is in the ground (equator) state, namely, 
$(\beta_1, \beta_2)=(1, 0)$ and $(\gam_1, \gam_2)=(1, e^{i\theta})/\sqrt{2}$. 
We observe that $\rho^{(\mrc)}_{\mra,ee}$ is independent of the phase $\theta$ 
of the initial photon qubit and increases in proportion to $|\alp|^2$, 
and that $\rho^{(\mrc)}_{\mra,eg}$ is an oscillating function of $\theta$
and its amplitude grows by increasing $|\alp|^2$.
These observations are in qualitative accordance 
with Eqs.~(\ref{eq:xiee}) and (\ref{eq:xieg}), 
which predicts that $\rho^{(\mrc)}_{\mra,ee}=|\alp|^2/2$ and 
$\rho^{(\mrc)}_{\mra,eg}=|\alp|^2 e^{i\theta}/2$.
These results indicate that the phase information of the initial photon qubit
is successfully transferred to the final atom qubit.

In Fig.~\ref{fig:PtoA}(c), we present the density matrix $\hat{\rho}^{(\mrs)}_\mra$
of the final atom qubit assuming the single-photon input, 
estimated by the aforementioned procedures. 
More details on the estimation are presented in Appendix~\ref{app:PtoA}.
The initial photon qubit is in one of the six cardinal states
[$|\om_L\ra$, $|\om_H\ra$, and $(|\om_L\ra+e^{in\pi/4}|\om_H\ra)/\sqrt{2}$ for $n=0,\cdots,3$].
The agreement between the initial photon and final atom qubits is fairly good 
and the averaged fidelity for the six cardinal states reaches 0.829.
The principal origin of the infidelity would be 
the short $T_1$ ($\sim 0.9~\mu$s) of the superconducting atom,  
which is comparable to the time required for the state tomography of the final atom qubit.
An exceptionally high fidelity is attained when the initial photon qubit is in $|\om_L\ra$ 
[first panel in Fig.~\ref{fig:PtoA}(c)].
This is because the atom remains in the ground state ($|g,0\ra \approx |\tone\ra$)
throughout the gate operation and is unaffected by the short $T_1$.

\subsection{State transfer: atom to photon}\label{ssec:3B}
\subsubsection{Procedures for density matrix estimation}\label{ssec:3B1}
Here, we demonstrate the atom-to-photon state transfer. 
More concretely, from the amplitudes of the final photon-qubit pulse 
(after reflection in Port~1) for the coherent-state input, 
we estimate the density matrix $\hat{\rho}_\mrp^{(\mrs)}$ 
of the final photon qubit assuming the single-photon input. 
The final amplitude $\xi(t)$ is given by $\xi(t)=\la \psi_f|\ha|\psi_f\ra$, 
where $\ha$ is the annihilation operator for a propagating photon in Port~1. 
Using Eq.~(\ref{eq:psif}), $\xi(t)$ is given, 
up to the first order in $|\alp|$, by
\bea
\xi(t) & = & 
\alp(|\beta_1|^2 \gam_1 + \beta_1\beta_2^* \gam_2)\psi_L(t) + 
\alp(\beta_1^*\beta_2 \gam_1 + |\beta_2|^2 \gam_2)\psi_H(t), 
\label{eq:xif}
\eea
where $\psi_{L(H)}(t)=\la 0|\ha|\om_{L(H)} \ra$ 
is the single-photon amplitude of the lower (higher) frequency component. 
When the initial photon-qubit pulse is monochromatic at $\om_L$, 
the final amplitude $\xi_L(t)$ is given,
by putting $(\gam_1, \gam_2)=(1, 0)$ in Eq.~(\ref{eq:xif}), by
\bea
\xi_L(t) & = & 
\alp |\beta_1|^2 \psi_L(t) + \alp \beta_1^*\beta_2 \psi_H(t). 
\label{eq:xiL}
\eea
This equation implies that
the initial monochromatic pulse may become dichromatic after reflection,
depending on the initial atomic state. 
However, when the atom is in the ground state initially, 
the final pulse remains monochromatic at $\om_L$.  
We denote its amplitude by $\zeta_L(t)$. 
Putting $(\beta_1, \beta_2)=(1, 0)$ in Eq.~(\ref{eq:xiL}), we have
\bea
\zeta_L(t) & = & \alp \psi_L(t). 
\label{eq:fL}
\eea
Similarly, when the initial pulse is monochromatic at $\om_H$, 
the final amplitude $\xi_H(t)$ is given by
\bea
\xi_H(t) & = & 
\alp \beta_1\beta_2^* \psi_L(t) + \alp |\beta_2|^2 \psi_H(t). 
\label{eq:xiH}
\eea
$\zeta_H(t)$ is the result for the initial atom being in the excited state. 
Putting $(\beta_1, \beta_2)=(0, 1)$ in Eq.~(\ref{eq:xiH}), we have
\bea
\zeta_H(t) & = & \alp \psi_H(t). 
\label{eq:fH}
\eea
We denote the overlap integral between $\zeta_i(t)$ and $\xi_j(t)$
($i,j=L,H$) by 
\bea
\eta_{ij} = \int dt \zeta_i^*(t) \xi_j(t). 
\label{eq:etaij}
\eea
Since $\psi_L(t)$ and $\psi_H(t)$ are orthogonal to each other
due to the different carrier frequencies, we obtain
$\eta_{LL}=C|\alp|^2|\beta_1|^2$, 
$\eta_{HH}=C|\alp|^2|\beta_2|^2$, 
$\eta_{LH}=C|\alp|^2 \beta_1 \beta_2^*$, and  
$\eta_{HL}=C|\alp|^2 \beta_1^* \beta_2$,   
where $C=\int dt |\psi_j(t)|^2$ ($j=L,H$).

On the other hand, 
our target quantity here is the density matrix $\hat{\rho}_\mrp^{(\mrs)}$ 
of the final photon qubit assuming the single-photon input. 
From the right-hand side of Eq.~(\ref{eq:swap2}), 
we immediately have
$\hat{\rho}_\mrp^{(\mrs)}=(\beta_1|\om_L\ra + \beta_2|\om_H\ra)(\beta_1^* \la\om_L| + \beta_2^*\la\om_H|)$. 
Therefore, the following 2$\times$2 matrix,
\bea
\hat{\eta} &=&
\frac{1}{\eta_{LL}+\eta_{HH}}
\left(\begin{array}{cc}
\eta_{LL} & \eta_{LH} \\
\eta_{HL} & \eta_{HH}
\end{array}\right), 
\label{eq:2by2}
\eea
is identical to the target density matrix $\hat{\rho}_\mrp^{(\mrs)}$ in principle.

Thus, we can construct the target density matrix $\hat{\rho}^\mathrm{(s)}_\mrp$
from the measured amplitudes of the final photon-qubit pulse
by the following procedures: 
Preliminarily, setting the initial atom-qubit state at $|g\ra$ ($|e\ra$), 
we apply a monochromatic photon-qubit pulse at $\om_L$ ($\om_H$) and 
measure the final amplitude $\zeta_L(t)$ [$\zeta_H(t)$]. 
Then, for an arbitrary initial atom-qubit state, 
we apply a monochromatic pulse at $\om_L$ ($\om_H$) and 
measure the final amplitude $\xi_L(t)$ [$\xi_H(t)$]. 
We construct a 2$\times$2 matrix $\hat{\eta}$ from the overlap integrals 
between these output amplitudes [Eqs.~(\ref{eq:etaij}) and (\ref{eq:2by2})]. 
$\hat{\eta}$ is identical to the target density matrix $\hat{\rho}_\mrp^{(\mrs)}$ in principle, 
but is non-Hermitian in practice (see Table~\ref{table:B}). 
We estimate a proper one by the protocol presented in Appendix~\ref{app:AtoP}.

\begin{figure}[h]
\begin{center}
\includegraphics[scale=1.0]{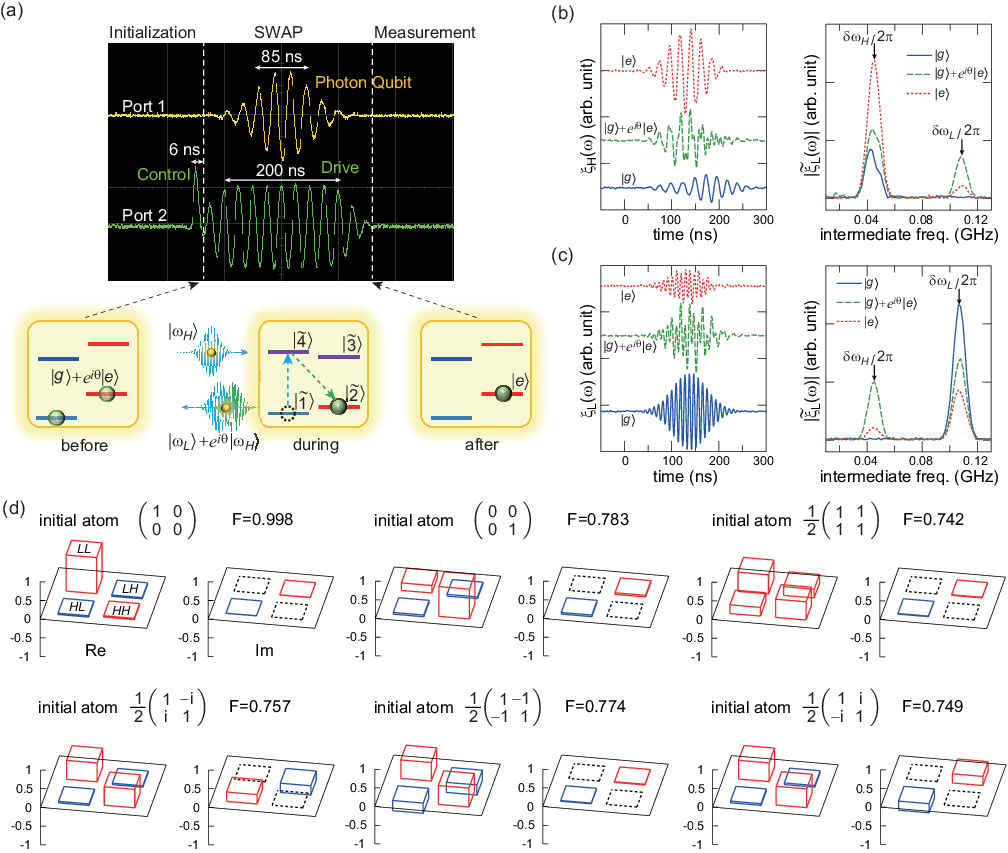}
\end{center}
\caption{
Atom-to-photon state transfer.
(a)~Pulse sequence depicted in the intermediate frequencies (IFs) 
and the energy diagrams before, during, and after the SWAP gate.
In the energy diagrams, the initial atom (photon) qubit is assumed to be 
in the equator (polar) state, namely, 
$(\beta_1, \beta_2)=(1, e^{i\theta})/\sqrt{2}$ and $(\gam_1, \gam_2)=(0, 1)$D
(b)~Measured amplitude of the output photon-qubit pulse 
for its initial frequency set at $\om_H$: 
time domain (left) and frequency domain (right).
The initial atom-qubit state (ground, equator, or excited) is indicated.
Since the four equator states yield similar spectra, only one of them is plotted. 
The mean photon number in the input pulse is $|\alp|^2=0.101$.
The time-domain amplitudes are measured by an analog-to-digital converter 
and are averaged over $1.5\times 10^4$ times.
(c)~The same plots as (b) for the initial frequency 
of the photon-qubit pulse set at $\om_L$.
(d)~Estimated density matrix $\hat{\rho}^{(\mrs)}_\mrp$
of the final photon qubit assuming the single-photon input. 
Positive (negative) values are indicated by red (blue) bars.
The fidelity to the initial atom qubit is indicated. 
}
\label{fig:AtoP}
\end{figure}
\subsubsection{Pulse sequence}
The pulse sequence to measure the amplitude $\xi(t)$ 
of the final photon-qubit pulse is shown in Fig.~\ref{fig:AtoP}(a). 
The measurement is composed of three steps.
(i)~Initialization: we wait for complete deexcitation of the atom.  
We then apply a control pulse with a Gaussian envelope 
at $\omega_{\rm ge}/2\pi = 5.835$~GHz from Port~2
to the atom to prepare it in one of the six cardinal states.
(ii)~SWAP gate: from Port~2, we apply a drive pulse 
with a flat-top envelope at $\omega_{\rm d}/2\pi = 5.775$~GHz to the atom 
to constitute an impedance-matched $\Lambda$ system [Fig.~\ref{fig:setup}(b)].
Within the pulse duration, we input from Port~1 
a weak monochromatic (at $\om_L/2\pi = 10.201$~GHz or $\om_H/2\pi = 10.263$~GHz) 
photon-qubit pulse with a Gaussian envelope. 
(iii)~Measurement: we measure the amplitude of 
the reflected photon-qubit pulse in Port~1,
which is dichromatic in general.
Note that $\om_L$ and $\om_H$ are slightly different 
from those in the ^^ ^^ photon-to-atom'' experiment, 
since the experiment was performed in the different cooling down of our dilution refrigerator 
(see Table~\ref{tab:params} 
for details on the experimental parameters).

All microwave pulses in Fig.~\ref{fig:AtoP}(a)
are generated by single-sideband modulation
and are shown in the intermediate frequencies (IFs),
similarly to those in Fig.~\ref{fig:PtoA}(a).
A carrier microwave at $\om_{\rm c}/2\pi = 10.308$~GHz and a Gaussian envelope 
with IF~$=\delta\om_H/2\pi = 0.045$~GHz ($\delta\om_L/2\pi = 0.107$~GHz) are mixed by an IQ mixer, 
obtaining the photon pulse with $\om_H = \om_{\rm c} - \delta\om_H$ ($\om_L = \om_{\rm c} - \delta\om_L$).
The final photon-qubit pulses after reflection 
by the atom are measured by an analog-digital converter 
after downconverted at $\om_{\rm c}$ in order to extract the signals at $\delta\om_H$ ($\delta\om_L$) 
as shown in the left panels in Figs.~\ref{fig:AtoP}(b) and (c).
The frequency-domain plots [right panels in Figs.~\ref{fig:AtoP}(b) and (c)] 
are obtained by applying the fast Fourier transform (FFT) to the time-domain plots.

\subsubsection{Results and discussions}
In Fig.~\ref{fig:AtoP}(b), 
we plot the final amplitudes of the photon-qubit pulse
in the time [$\xi_H(t)$, Eq.~(\ref{eq:xiH})]
and frequency [$\widetilde{\xi}_H(\om)$, Fourier transform of $\xi_H(t)$] domains, 
fixing its initial frequency at $\om_H$ and 
varying the initial atomic state.
Predictions by Eq.~(\ref{eq:xiH}) are as follows.
(i)~Putting $(\beta_1, \beta_2)=(0, 1)$, we have $\xi_H(t)=\alp\psi_H(t)$. 
Namely, when the atom is in the excited state initially, 
the final amplitude is monochromatic at $\om_H$,  
unchanged from the initial one. 
(ii)~Putting $(\beta_1, \beta_2)=(1, e^{i\theta})/\sqrt{2}$, 
we have $\xi_H(t)=(\alp/2)[e^{-i\theta}\psi_L(t)+\psi_H(t)]$. 
Namely, when the atom is in the equator state initially, 
the final amplitude is dichromatic at $\om_L$ and $\om_H$ with equal magnitudes.
(iii)~Putting $(\beta_1, \beta_2)=(1, 0)$, we have $\xi_H(t)=0$. 
Namely, when the atom is in the ground state initially,  
the final amplitude vanishes. 
We observe that the measured amplitudes in Fig.~\ref{fig:AtoP}(b)
are in qualitative agreement with these predictions. 
However, we also find discrepancies from these predictions, 
such as non-vanishing signal at $\om_H$ for the initial atom in $|g\ra$  
and appearance of the $\om_L$ component for the initial atom in $|e\ra$. 
We attribute the principal reason for the former discrepancy
to the imperfect constitution of an impedance-matched $\Lambda$ system
[namely, difference in the $|\tfour\ra \to |\tone\ra$ and $|\tfour\ra \to |\ttwo\ra$ 
decay rates in Fig.~\ref{fig:lvs}(b)]
and the latter to the imperfect initialization of the atom qubit,
both of which originate in the fluctuation in the transition frequency 
$\om_\mathrm{ge}$ of the superconducting atom.
Note that the drastic attenuation of the final amplitude 
in (iii) does not mean the decrease of the reflected photon in Port~1: 
the input photon is mostly downconverted to $\om_L$
but its amplitude is unobservable in this experiment due to the inelasticity of scattering. 
This quantum process (single-photon Raman interaction~\cite{dop_th_1,dop_th_2,dop_th_3,our_1,our_2,dayan1,dayan2,dayan3})
has been applied for the single microwave photon detection~\cite{spd1,spd2}. 
Figure~\ref{fig:AtoP}(c) shows the results for 
the initial frequency of the photon-qubit pulse tuned at $\om_L$.
The results are contrastive to those in Fig.~\ref{fig:AtoP}(b)
and are in qualitative accordance with Eq.~(\ref{eq:xiL}).

In Fig.~\ref{fig:AtoP}(d), we present the density matrix $\hat{\rho}^{(\mrs)}_\mrp$ 
of the final photon qubit assuming the single-photon input, 
estimated by the aforementioned procedures. 
More details on estimation are presented in Appendix~\ref{app:AtoP}.
The fidelity to the initial atom qubit,
which is prepared to be in one of the six cardinal states,
is fairly good and the averaged fidelities reaches 0.801. 
An exceptionally high fidelity is attained 
when the initial atom is in $|g\ra$ [first panel in Fig.~\ref{fig:AtoP}(c)], 
because the atom remains in the ground state throughout the gate operation
and is unaffected by the short $T_1(\sim 0.9~\mu$s) of the superconducting atom.
We observe that, when the initial atom qubit is in the equator states, 
the fidelities become substantially lower than 
those for the photon-to-atom state transfer [Fig.~\ref{fig:PtoA}(c)].
Presently, we do not fully understand the reason for that.
One possible reason might be that
the amplitude of the frequency-converted component
[$\om_{L/H}$ component in the right panel of Fig.~\ref{fig:AtoP}(b/c)],
which becomes observable only when the initial atom
is in a superposition state, 
is more fragile against the pure dephasing
than the amplitude of the unconverted component.

\section{conclusion}\label{sec:sum}
We have demonstrated a deterministic SWAP gate 
between a superconducting qubit and a frequency-encoded
microwave-photon qubit. 
More concretely, we have confirmed the bidirectional (photon-to-atom and atom-to-photon) transfer of the qubit state.
The photon qubit for this gate is a single-photon pulse propagating in a waveguide, 
but we used a weak coherent-state pulse instead for demonstration.

To confirm the photon-to-atom qubit transfer, 
we applied a monochromatic or dichromatic photon-qubit pulse,  
which corresponds to one of the six cardinal states of the photon qubit, 
to the dressed atom-resonator coupled system (impedance-matched $\Lambda$ system). 
After reflection of this pulse, 
we performed a state tomography of the final atom qubit. 
From the dependencies of the density matrix elements 
on the mean input photon number, 
we constructed the density matrix of the final atom qubit 
assuming the single-photon input. 
The fidelity to the initial photon qubit reaches 0.829 on average.
On the other hand, to confirm the atom-to-photon qubit transfer, 
we prepared the initial atom qubit to be in one of the six cardinal states
and applied a monochromatic photon-qubit pulse to the $\Lambda$ system. 
From the measured amplitudes of the final photon-qubit pulse, 
we constructed the density matrix of the final photon qubit 
assuming the single-photon input.
The fidelity to the initial atom qubit reaches 0.801 on average.

Although the fidelities of the qubit-state transfer here
are still insufficient for practical application, 
the principal reason for these infidelities is 
the short lifetime of the superconducting atom, 
which can readily be overcome with the current qubit fabrication technology. 
We hope that the present scheme for the atom-photon SWAP gate, 
equipped with distinct merits such as 
simplicity of the setup and {\it in-situ} gate tunability, 
would help the distributed quantum computation with superconducting qubits in near future.

\section*{Acknowledgments}
The authors are grateful to T. Shitara, S. Masuda, and A. Noguchi for fruitful discussions.
This work was supported by JST Moonshot R\&D (JPMJMS2061-2-1-2, JPMJMS2062-10, JPMJMS2067-3), 
JSPS KAKENHI (22K03494) and JST PRESTO (JPMJPR1761).

\appendix
\section{Density matrix estimation: photon to atom}
\label{app:PtoA}
Here, we present the details on the density matrix estimation 
for the photon-to-atom state transfer [Fig.~\ref{fig:PtoA}(c)]. 

\subsection{State tomography of $\hat{\rho}^{(\mrc)}_{\mra}$}
We first discuss how to determine
the density matrix $\hat{\rho}^{(\mrc)}_{\mra}$ 
of the final atom qubit from the measurement data. 
In the tomography stage of Fig.~\ref{fig:PtoA}(a), 
we first perform one of the six kinds of one-qubit gates to the atom.
The unitary matrices corresponding to these gates are 
\bea
\hat{U}_j &=& 
\begin{cases}
\hat{I} & (j=1) \\
\hsig_x & (j=2) \\
(\hat{I}-i\cos(\frac{j\pi}{2})\hsig_x-i\sin(\frac{j\pi}{2})\hsig_y)/\sqrt{2} & (j=3,\cdots,6)
\end{cases}.
\label{eq:U1to6}
\eea
We measure the excitation probability of the atom by dispersive readout.
We denote the {\it measured} probability after $j$th gate by $\widetilde{p}_j$. 
We estimate the atomic density matrix $\hat{\rho}^{(\mrc)}_{\mra}$ 
from the measurement data set $(\widetilde{p}_1,\cdots,\widetilde{p}_6)$.

We parameterize $\hat{\rho}^{(\mrc)}_{\mra}$ as 
\bea
\hat{\rho}^{(\mrc)}_{\mra} &=& 
(\hat{I} + a_x \hsig_x + a_y \hsig_y + a_z \hsig_z)/2,
\label{eq:hrhoc}
\eea
where $a_x$, $a_y$, and $a_z$ are the Bloch vector components,  
which are real and satisfy $a_x^2+a_y^2+a_z^2 \leq 1$. 
With this density matrix, the {\it expected} excitation probability 
$p_j$ after $j$th gate is given by 
$p_j=\mathrm{Tr}\{ (\hat{1}+\hat{\sig}_z) \hat{U}_j \hat{\rho}^{(\mrc)}_{\mra} \hat{U}_j^{\dagger} \}/2$. 
Using Eqs.~(\ref{eq:U1to6}) and (\ref{eq:hrhoc}), we have
$p_1=(1+a_z)/2$, $p_2=(1-a_z)/2$, $p_3=(1+a_x)/2$, 
$p_4=(1+a_y)/2$, $p_5=(1-a_x)/2$, and $p_6=(1-a_y)/2$. 
We determine the parameters $a_x$, $a_y$, and $a_z$ 
so as to minimize the sum of squared errors, 
\bea
S(a_x,a_y,a_z) 
&=& \sum_{j=1}^6 (p_j-\widetilde{p}_j)^2.
\eea
This is rewritten as
$S(a_x,a_y,a_z)=(a_x-\overline{a}_x)^2 + (a_y-\overline{a}_y)^2 + (a_z-\overline{a}_z)^2 + \cdots$, 
where
\bea
\overline{a}_x &=& \widetilde{p}_3-\widetilde{p}_5,
\\
\overline{a}_y &=& \widetilde{p}_4-\widetilde{p}_6,
\\
\overline{a}_z &=& \widetilde{p}_1-\widetilde{p}_2.
\eea
%
%
Therefore, if the point P$(\overline{a}_x, \overline{a}_y, \overline{a}_z)$ 
is inside of the unit sphere, $S$ is minimized at this point. 
In contrast, if the point P is out of the unit sphere, 
$S$ is minimized at the projection of point~P to the unit-sphere surface in the radial direction.
%
%
Therefore, 
\bea
(a_x,a_y,a_z) &=& \frac{(\overline{a}_x, \overline{a}_y, \overline{a}_z)}
{\max\left(1,\sqrt{\overline{a}_x^2+\overline{a}_y^2+\overline{a}_z^2}\right)}.
\eea
In Table~\ref{table:A}, 
setting the initial photon qubit at $|\psi_{\mrp}\ra=(|\om_L\ra-i|\om_H\ra)/\sqrt{2}$ for example, 
we present the measurement data set $(\widetilde{p}_1,\cdots,\widetilde{p}_6)$ and 
the estimated Bloch vector components
for various input photon number $|\alp|^2$. 

\begin{table}[h]
\caption{
Measured excitation probabilities $\widetilde{p}_j$ and 
the estimated Bloch vector components $(a_x, a_y, a_z)$
for various input photon number $|\alp|^2$. 
The initial photon-qubit state is $|\psi_{\mrp}\ra=(|\om_L\ra-i|\om_H\ra)/\sqrt{2}$. 
}
\label{table:A}
\centering
\begin{tabular}{|r|rrrrrr|rrr|}
\hline
\multicolumn{1}{|c|}{$|\alp|^2$} & 
\multicolumn{1}{c}{$\widetilde{p}_1$} & 
\multicolumn{1}{c}{$\widetilde{p}_2$} & 
\multicolumn{1}{c}{$\widetilde{p}_3$} & 
\multicolumn{1}{c}{$\widetilde{p}_4$} & 
\multicolumn{1}{c}{$\widetilde{p}_5$} & 
\multicolumn{1}{c|}{$\widetilde{p}_6$} & 
\multicolumn{1}{c}{$a_x$} & 
\multicolumn{1}{c}{$a_y$} & 
\multicolumn{1}{c|}{$a_z$} \\ 
\hline
$0.0000$ & $0.0371$ & $0.9226$ & $0.4953$ & $0.4965$ & $0.4812$ & $0.4821$ & 
$0.0141$ & $0.0144$ & $-0.8855$
\\
$0.1950$ & $0.0721$ & $0.8306$ & $0.4795$ & $0.5452$ & $0.4823$ & $0.4124$ & 
$-0.0028$ & $0.1328$ & $-0.7585$
\\
$0.4992$ & $0.1950$ & $0.7358$ & $0.4800$ & $0.6442$ & $0.4697$ & $0.3084$ &
$0.0103$ & $0.3358$ & $-0.5408$
\\
\hline
\end{tabular}
\end{table}

\subsection{Estimation of $\hat{\rho}^{(\mrs)}_{\mra}$ from $\hat{\rho}^{(\mrc)}_{\mra}$}
Here, we discuss how to estimate the atomic density matrix $\hat{\rho}^{(\mrs)}_{\mra}$ 
assuming the single-photon input from the one $\hat{\rho}^{(\mrc)}_{\mra}$
for the coherent-state input. 
Similarly to Eq.~(\ref{eq:hrhoc}), we parametrize the target density matrix as
\bea
\hat{\rho}^{(\mrs)}_{\mra} &=& (\hat{I} + b_x\hsig_x + b_y\hsig_y + b_z\hsig_z)/2. 
\label{eq:hrhos}
\eea
Then, From Eqs.~(\ref{eq:drhoda1}) and (\ref{eq:drhoda2}), we obtain
\bea
b_x &=& \frac{da_x}{d|\alp|^2}, 
\\
b_y &=& \frac{da_y}{d|\alp|^2}, 
\\
b_z &=& \frac{da_z}{d|\alp|^2}-1. 
\eea
Therefore, we can estimate $b_x$ from the dependence of $a_x$ 
on the mean input photon number $|\alp|^2$. 
%
%
Assuming a linear dependence and employing the least square method, 
we determine the slope $\overline{b}_x$
from the data set of $\{ |\alp|^{2(j)}, a_x^{(j)} \}$ for $j=1,\cdots,N$, 
where $N$ is the number of the data set.
$\overline{b}_x$ is then given by
\bea
\overline{b}_x &=& \frac{NC_2-C_3C_4}{NC_1-C_3^2},
\eea
where 
$C_1=\sum_j (|\alp|^{2(j)})^2$, 
$C_2=\sum_j |\alp|^{2(j)} a_x^{(j)}$, 
$C_3=\sum_j |\alp|^{2(j)}$, and
$C_4=\sum_j a_x^{(j)}$. 
$\overline{b}_y$ and $\overline{b}_z$ are determined similarly. 
If the point Q$(\overline{b}_x, \overline{b}_y, \overline{b}_z)$ is outside of the unit sphere, 
we project this point to the unit-sphere surface in the radial direction. 
Therefore, 
\bea
(b_x,b_y,b_z) &=& \frac{(\overline{b}_x, \overline{b}_y, \overline{b}_z)}
{\max\left(1,\sqrt{\overline{b}_x^2+\overline{b}_y^2+\overline{b}_z^2}\right)}.
\eea

From the data set presented in Table~\ref{table:A}, 
we have $b_x=-0.0032$, $b_y=0.6459$, and $b_z=-0.3074$. 
Accordingly, $\rho^{(\mrs)}_{\mra,ee}=(1+b_z)/2=0.3463$ 
and $\rho^{(\mrs)}_{\mra,eg}=(b_x+ib_y)/2=-0.0016-0.3229i$
[the last panel in Fig.~\ref{fig:PtoA}(c)]. 

\section{Density matrix estimation: atom to photon}\label{app:AtoP}
Here, we present the details on the density matrix estimation 
for the atom-to-photon state transfer [Fig.~\ref{fig:AtoP}(d)]. 

\subsection{Estimation of $\hat{\rho}^{(\mrs)}_{\mrp}$ from $\hat{\eta}$}
According to the arguments in Sec.~\ref{ssec:3B}, 
the matrix $\hat{\eta}$ constructed directly from the experimental data 
[Eqs.~(\ref{eq:etaij}) and (\ref{eq:2by2})] is, in principle, 
identical to the target density matrix $\hat{\rho}^{(\mrs)}_{\mrp}$.
However, as we observe in Table~\ref{table:B}, $\hat{\eta}$ is non-Hermite in practice. 
We therefore estimate a proper density matrix $\hat{\rho}^{(\mrs)}_{\mrp}$
from $\hat{\eta}$ by the following procedures. 

Similarly to Eq.~(\ref{eq:hrhoc}), we parameterize the proper density matrix as
\bea
\hat{\rho}^{(\mrs)}_{\mrp} &=& (\hat{I} + c_x \hsig_x + c_y \hsig_y + c_z \hsig_z)/2,
\label{eq:hrhosp}
\eea
where $c_x$, $c_y$, and $c_z$ are real and satisfy $c_x^2 + c_y^2 + c_z^2 \leq 1$. 
We choose $c_x$, $c_y$ and $c_z$ so as to minimize the distance $L$ 
between $\hat{\eta}$ and $\hat{\rho}^{(\mrs)}_{\mrp}$, 
which we quantify by 
\bea
L(c_x,c_y,c_z) &=& \sum_{j=x,y,z} |\la\hat{\sig}_j\ra_{\eta} - \la\hat{\sig}_j\ra_{\rho}|^2, 
\label{eq:Lxyz}
\eea
where $\la\hat{\sig}_j\ra_{\eta}=\mathrm{Tr}\{\hat{\sig}_j \hat{\eta}\}$
and $\la\hat{\sig}_j\ra_{\rho}=\mathrm{Tr}\{\hat{\sig}_j \hat{\rho}^{(\mrs)}_{\mrp}\}$. 
Since $\la\hat{\sig}_x\ra_{\eta}=\eta_{LH}+\eta_{HL}$, $\la\hat{\sig}_y\ra_{\eta}=i(\eta_{LH}-\eta_{HL})$, 
$\la\hat{\sig}_z\ra_{\eta}=\eta_{LL}-\eta_{HH}$, $\la\hat{\sig}_x\ra_{\rho}=c_x$, 
$\la\hat{\sig}_y\ra_{\rho}=c_y$, and $\la\hat{\sig}_z\ra_{\rho}=c_z$, 
Eq.~(\ref{eq:Lxyz}) is rewritten as
$L(x,y,z)=(c_x-\overline{c}_x)^2 + (c_y-\overline{c}_y)^2 + (c_z-\overline{c}_z)^2 + \cdots$, 
where
\bea
\overline{c}_x &=& \mathrm{Re}(\eta_{LH}+\eta_{HL}), 
\\
\overline{c}_y &=& \mathrm{Im}(\eta_{HL}-\eta_{LH}), 
\\
\overline{c}_z &=& \mathrm{Re}(\eta_{LL}-\eta_{HH}).  
\eea
Therefore, if the point R$(\overline{c}_x, \overline{c}_y, \overline{c}_z)$
is inside of the unit sphere, $L$ is minimized at this point.
On the other hand, if the point~R is out of the unit sphere, 
$L$ is minimized at the projection of point~R to the unit sphere.
Therefore, 
\bea
(c_x,c_y,c_z) &=& \frac{(\overline{c}_x, \overline{c}_y, \overline{c}_z)}{\max\left(1,\sqrt{\overline{c}_x^2+\overline{c}_y^2+\overline{c}_z^2}\right)}. 
\eea

In Table~\ref{table:B}, 
we present the matrix elements of $\hat{\eta}$ and $\hat{\rho}^{(\mrs)}_{\mrp}$
for various input photon number $|\alp|^2$.
The initial atom-qubit state is chosen as $|\psi_{\mra}\ra=(|g\ra-i|e\ra)/\sqrt{2}$ for example, 
and the fidelity is that between the initial atom and final photon qubits,
$F=\la \psi_\mra|\hat{\rho}^{(\mrs)}_{\mrp}|\psi_\mra \ra$.
We observe that the estimated density matrix is 
mostly insensitive to the input photon number $|\alp|^2$. 
In Fig.~\ref{fig:AtoP}(c), 
we employ the averaged density matrix over the four cases as $\hat{\rho}^{(\mrs)}_{\mrp}$.

\begin{table}[h]
\caption{
Elements of density matrices $\hat{\eta}$ 
and $\hat{\rho}^{(\mrs)}_{\mrp}$. 
The initial atom-qubit state is $|\psi_{\mra}\ra=(|g\ra-i|e\ra)/\sqrt{2}$ for example
and the mean input photon number $|\alp|^2$ is varied.
$\rho^{(\mrs)}_{\mrp,HH}(=1-\rho^{(\mrs)}_{\mrp,LL})$ and 
$\rho^{(\mrs)}_{\mrp,HL}(=\rho^{(\mrs)*}_{\mrp,LH})$ are not shown. 
}
\label{table:B}
\centering
\begin{tabular}{|c|cccc|cc|c|}
\hline
$|\alp|^2$ & $\eta_{LL}$ & $\eta_{LH}$ & $\eta_{HL}$ & $\eta_{HH}$ 
& $\rho^{(\mrs)}_{\mrp,HH}$ & $\rho^{(\mrs)}_{\mrp,HL}$ & fidelity \\
\hline
$0.048$ & $0.515+0.057i$ & $-0.095+0.187i$ & $-0.003-0.277i$ & $0.485-0.057i$ & $0.485$ & $-0.049-0.232i$ & $0.732$ \\
$0.101$ & $0.509+0.043i$ & $-0.113+0.215i$ &  $0.042-0.333i$ & $0.491-0.043i$ & $0.491$ & $-0.036-0.274i$ & $0.774$ \\
$0.163$ & $0.502+0.056i$ & $-0.106+0.204i$ &  $0.030-0.282i$ & $0.498-0.056i$ & $0.498$ & $-0.038-0.243i$ & $0.743$ \\
$0.206$ & $0.503+0.058i$ & $-0.121+0.162i$ &  $0.038-0.330i$ & $0.497-0.058i$ & $0.497$ & $-0.042-0.246i$ & $0.746$ \\
\hline
average & & & & & $0.493$ & $-0.041-0.249i$ & $0.749$ \\
%
\hline
\end{tabular}
\end{table}

\section{Experimental information}\label{app:exp}
\subsection{Experimental setup}\label{app:exp1}
\begin{figure}[h]
\begin{center}
\includegraphics[width=100mm]{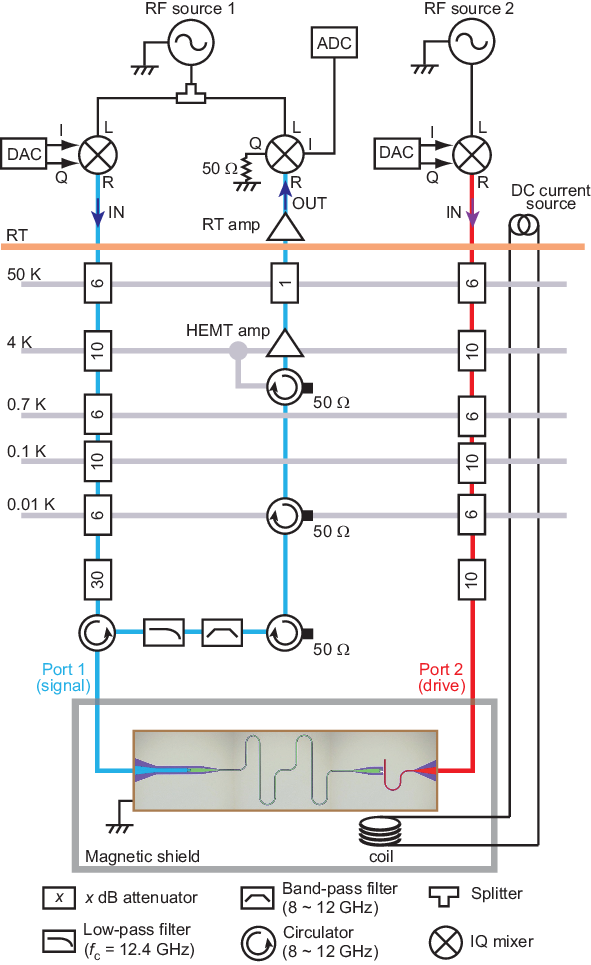}
\end{center}
\caption{Schematic of the experimental setup.}
\label{fig:Eset}
\end{figure}
Figure~\ref{fig:Eset} shows a schematic of the measurement setup 
composed of room-temperature microwave instruments and low-temperature wirings 
with microwave components in a cryogen-free $^3$He/$^4$He dilution refrigerator.

The photon-qubit 
(drive) pulses applied to Port~$1$ ($2$) are 
generated by mixing the continuous microwave from the RF source~$1$ ($2$) 
with pulses that have an intermediate frequency (IF) generated by a DAC (digital to analog converter) with a sampling rate of $1$~GHz.
The Gaussian pulses are used for the photon-qubit pulses, 
while the flat-top pulses, in which the rising and falling edges of the pulse envelope 
are smoothed by a Gaussian function with full width at half maximum of $40$~ns 
in its voltage amplitude, are employed for the drive pulses.

The signal pulses are heavily attenuated by a series of attenuators 
implemented in the input microwave semi-rigid cable with a total attenuation of $68$~dB 
and applied to the $\lambda/2$ resonator through a circulator to separate the input and reflected signal.
The reflected signal is led to a cryogenic HEMT amplifier 
mounted at a $4$~K stage of the dilution refrigerator via low-pass and band-pass filters 
and three circulators with 50 $\Omega$ terminations and amplified by $\sim 38$~dB 
followed by further amplification with a room-temperature low-noise amplifier, whose gain is $\sim 33$~dB.
The output signal is down-converted to IF at an IQ mixer 
using a continuous microwave split the same signal used for the photon-qubit pulse generation.
The I component of the reflected signals is sampled at $1.6$~GHz/s by an ADC (analog to digital converter).

The drive pulses are applied through the input microwave semi-rigid cable with a total attenuation of $48$~dB to control the states of an atom.

\subsection{Device and parameters}\label{app:exp2}
A device employed in our experiments is composed of a  $\lambda/2$ superconducting coplanar waveguide resonator 
and a superconducting flux qubit containing three Josephson junctions [Fig.~\ref{fig:setup}(b)].
They are coupled capacitively and operated in the dispersive regime.
We adopted the same design and fabrication processes for the device described in Ref.~\cite{spd2,our_2}.

In Table~\ref{tab:params}, we summarize the system
parameters for the SWAP experiments described in the main text.
Since each SWAP experiment was performed in the different cooling down of our dilution refrigerator,
$\omega_{\rm ge}$ and related parameters ($\chi$, $\om_\mathrm{d}$, $\om_L$ and $\om_H$) changed slightly.
The other parameters ($\om_r$ and $\kap$)
are independent of $\omega_{\rm ge}$ and remain unchanged.

\begin{table*}[h]
\caption{Device parameters for the ``photon to atom" and the ``atom to photon" SWAP experiments (unit: GHz).
}
\label{tab:params}
\centering
\begin{tabular}{ccc}
\hline
~Parameters~ &~Photon $\rightarrow$ Atom~ &~Atom $\rightarrow$ Photon~ \\
\hline \hline
$\om_{\rm r}/2\pi$&10.258 &10.258\\
$\om_{\rm ge}/2\pi$&5.839 &5.835\\
$2\chi/2\pi$&0.073 &0.073\\
$\kap/2\pi$ & 0.024 & 0.024 \\
\hline 
$\om_{\rm d}/2\pi$&5.785 &5.775\\
$\om_L/2\pi$ &10.208 &10.201\\
$\om_H/2\pi$&10.266 &10.263\\
\hline
\end{tabular}
\end{table*}


\end{document}